# Instantaneous Frequency representation used for CPA laser simulation

THOMAS OKSENHENDLER[1*], STEFAN BOCK[2], ULRICH SCHRAMM[2],

[1]*iTEOX, 14 avenue Jean Jaurès, 91940 Gometz-le-chatel, France*
[2]*Helmholtz-Zentrum Dresden-Rossendorf (HZDR), Bautzner Landstr. 400, 01328 Dresden, Germany*

*Corresponding author: thomas@iteox.com

We present a novel intuitive graphical method for the simulation of non-linear effects on stretched pulses characterized by a large time-bandwidth product. By way of example, it allows precise determination of effects occurring in CPA (chirped pulse amplification) laser chains, such as the pre-pulse generation by the non-linear Kerr effect. This method is not limited to first order dispersion and can handle all resulting distortions of the generated pre-pulse.

## I. Introduction

Ultra-intense laser systems are used for a variety of applications in the fields of high energy density physics and relativistic laser plasma physics [1]. Achievable contrast ratios are in the range of $10^{-12}$ on temporal scales between ns and few 10 ps prior to the main pulse. Amplification of stretched pulses (i.e., chirped pulse amplification, CPA [2]) is conventionally performed in regenerative and multipass amplifiers. Despite its capacity to lower nonlinearities in the amplifiers, the CPA technique when used for extremely high-power system (PW range) still accumulates nonlinear distortions due to the Kerr effect characterized by the maximum value of the B-integral. The B-integral represents the nonlinear temporal phase shift acquired after propagation through the system. Typical levels of the B-integral are in the range of 0.1 to 1 [3,4,5]. To simulate the consequence of this nonlinearity on the temporal profile of the output beam it requires to simulate nonlinear effects on stretched pulses characterized by a large time-bandwidth product. As an example, recent observations [6] have shown that the temporal characteristics of a pre-pulse may differ significantly from the main and post pulses from which it was generated via temporal diffraction.

These observations, significantly affecting the interaction of high-power laser pulses with matter, have renewed the interest in intuitive simulation capabilities of complex pulse shapes developing through the CPA process.

In classical finite-difference time-domain or split-step Fourier methods large time-bandwidth product laser pulses like the ones used in CPA are difficult to handle. Although these methods solve the pulse propagation problem, they are time and memory consuming as the temporal resolution has to be in the order of the Fourier transform limited pulse width while the excursion should remain larger than the pulse width. For typical CPA conditions this requires $10^4$ or $10^5$ points at least.

For the sake of simplicity, we restrict our attention to a one dimensional model for a scalar field and use Maxwell's wave equation in the form [6]

$$\frac{\partial^2 E}{\partial z^2} - \frac{1}{c^2}\frac{\partial^2 D}{\partial \tau^2} = 0, \qquad (1)$$

where $E = E(\tau,z)$ is the electric field, D is the electric displacement and c the velocity of light in vacuum. The constitutive relation between D and E takes into account both dispersion and nonlinearity of the medium

$$D = \varepsilon E + P_{NL}, \qquad (2)$$

where ε is the dielectric permittivity ($\varepsilon = n^2$ with n being the refractive index), and $P_{NL}$ the nonlinear polarization. In our case, we will only consider the third order nonlinear polarization enabling effects like four wave mixing, self-phase modulation and cross phase modulation: $P_{NL} = \chi^{(3)}|E|^2 E$, where $\chi^{(3)}$ is the third order non linear susceptibility.

The first term of the electric displacement includes all linear terms and has a simple form in the frequency domain:

$$\tilde{E}_{out}(\omega) = \tilde{H}(\omega)\tilde{E}_{in}(\omega), \qquad (3)$$

where $\tilde{E}_{out}(\omega)$ and $\tilde{E}_{in}(\omega)$ are the Fourier transforms of the output and input signals, $\omega = 2\pi f$ is the optical pulsation where f is the optical frequency. The frequency response function $\tilde{H}(\omega)$ includes the spectral transmission $T(\omega)$ as its amplitude and the spectral phase dispersion $\varphi(\omega)$ as its argument, $\tilde{H}(\omega) = T(\omega)e^{i\varphi(\omega)}$. The simulation of this term is straight forward in the spectral domain.

But as for the non-linear Schrodinger Equation, the non-linear term requires to be simulated in the temporal domain [8,9], a Fast Fourier transform is usually used to pass from spectral to temporal domain and vice-versa. As already mentioned, the huge temporal excursion due to the stretching ratio of the CPA will introduce a huge number of points that tends to oversize the memory of standard computers.

In this paper we propose an alternative method that keeps advantages of both domains without needing any direct relation between spectral and temporal resolution and excursion. It also has the advantage of being intuitive as a graphical representation of stretched laser pulses with huge time-bandwidth products.

## I. Instantaneous Frequency representation for a sample CPA simulation

Large time-bandwidth product pulses are difficult to represent. In the general case, there is no other option than to use a very large temporal excursion that covers the full duration and a large bandwidth that covers the spectral amplitude of the pulse. As both domains are linked by Fourier transformation, the temporal resolution is inversely proportional to the spectral excursion and vice-versa. To visualize such relation and large time-bandwidth product pulses, time-frequency representations like Wigner-Ville or spectrogram are commonly used. We here introduce the instantaneous frequency representation (IFR). The instantaneous frequency $\omega(\tau) = \partial\varphi/\partial\tau$ weighted by the associated spectral amplitude is fully representative for a laser pulse. To further increase the feeling of the physics meaning, the line (in the temporal domain) can be convoluted by the Fourier transform pulse profile temporal amplitude. The representation of a Fourier transform limited pulse is shown in Fig.1. For a Wigner-Ville representation, both domains would have the same number of points to easily calculate the Fourier Transform as represented by the grey grid in fig.1. Thus, the number of points N, the resolutions ($\delta f$, $\delta\tau$) and excursions ($\Delta F$, $\Delta T$) of both domains are fully determined:

$$\Delta F \delta\tau = \Delta T \delta f = 1 \text{ and } \frac{\Delta F}{\delta f} = \frac{\Delta T}{\delta\tau} = N = \frac{1}{\delta f \delta\tau}. \qquad (4)$$

For Fourier transform limited pulses, the time-bandwidth product is minimal and close to 0.5

$$\Delta\omega_{pulse}\Delta t_{pulse} \geq 0.5 \qquad (5)$$

where $\Delta\omega_{pulse}$ and $\Delta t_{pulse}$ are statistical widths (root mean squares).

On the IFR, as for Wigner-Ville distribution, the spectral or temporal amplitude can be recovered by projection of the curve along one dimension (Fig.1.1&2). A Fourier limited pulse is represented as a vertical line. The instantaneous frequency curve ($\tau(\omega) = (\partial\varphi/\partial\tau)_{\omega_i - \omega_0} = 0$) is represented by the black dotted line in Fig 1.3. Applying linear filters like spectral transmission or dispersion to the pulse can be done directly on the IFR with line-by-line modification. A pure spectral amplitude corresponds to a simple multiplication of the line $\omega_i$ by the spectral amplitude T($\omega_i$). After line-by-line multiplication, the new spectral amplitude is used to calculate the new temporal Fourier limit. This Fourier limit temporal profile is then applied onto the temporal pulse by convolution.

In this case, the constraint of having the same number of points and the temporal excursion limited by the spectral resolution is not limiting as the pulse time-bandwidth product is still minimal.

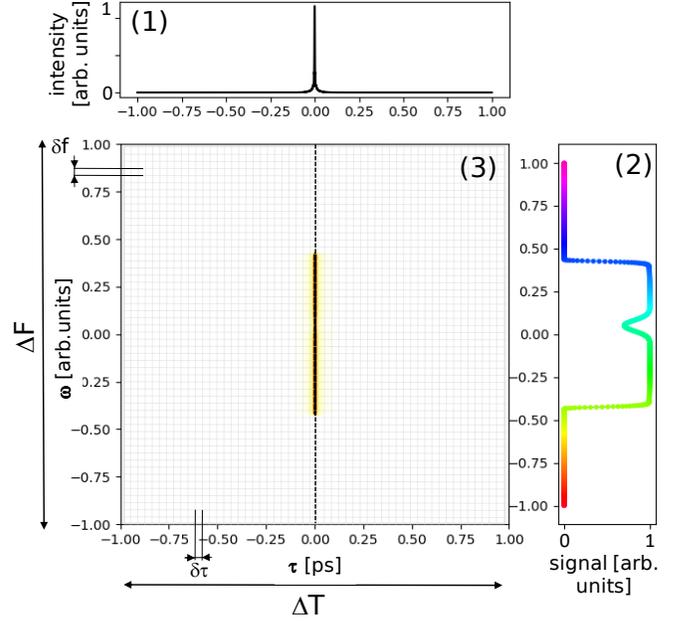

Fig. 1. Instantaneous frequency representation of a Fourier limited pulse in (1) temporal intensity, (2) spectral intensity, (3) instantaneous frequency with typically resolutions required in Fourier transformations (grid) for comparison. To avoid symmetric spectral shape while keeping a good visibility on the IFR, the pulse chosen exhibits a dip on its spectral intensity.

But in a CPA laser, the pulse is stretched. Its time-bandwidth product rises to about $10^5$. The temporal excursion of the pulse is then greater than that used for the simulation as shown by the grey grid in Fig.2.a.3. By using Fourier transform like in the Wigner-Ville or spectrogram simulation, the temporal domain is restricted to this area. The simulation then presents temporal aliasing harmful to the final result as shown in Fig.2.a.3 and by the blue curve in Fig.2.a.1. For conventional large-scale CPA lasers, the initial temporal domain is so small that it becomes nearly invisible on the IFR (Fig.2.b.3). With conventional Fourier transform, to avoid this temporal aliasing the only solution is to increase the number of samples proportionally on both axes.

This makes it difficult to use Fourier transformations for very strongly stretched pulses with huge time-bandwidth products, since the required number of points grows quadratically with the stretching rate.

While by using IFR, it is possible to stretch the temporal scale without restriction (fig.2.a&b). The stretching corresponds to a translation of the points of the instantaneous curve depending on the frequency. For each $\omega_i$ the position of the new point is modified by the relative delay from the central pulsation

$$d\tau(\omega_i) = (\partial\varphi/\partial\omega)_{\omega_i - \omega_0}. \qquad (6)$$

The operation has no temporal limitation and does not need a regular temporal pattern. It is a pure spectral operation line by line. The output temporal scale is not anymore linked to the spectral one by a regular point to point pattern. The temporal amplitude estimation will then require either to resample the curve on a regular temporal pattern or to consider the irregular pattern in the amplitude calculation. The IFR is then compatible to any time-bandwidth product.

Pulse replicas are also due to linear operations but combine both spectral amplitude and spectral phase. As an example a post-pulse delayed by $t_d$ is simply introduced by replicating the IF curve and translating the full curve by $t_d$. The absolute phase difference at $\omega_0$ between the pulses is ignored here. It can be considered by using an additional array with the phase values or by using complex weights rather than purely real ones. The calculation of the spectral amplitude is then more complex as it requires integrating of all different $\omega_i$ components. The sum of these components includes the phase terms that are frequency dependent $\varphi(\omega_i) = \omega_i t_d$. The spectral amplitude is then obtained through the integration

$$\tilde{A}(\omega_i) = \int_{-T}^{T} A(\tau)e^{i\omega_i \tau}d\tau. \qquad (7)$$

By symmetry, the temporal intensity is also recovered by integration as

$$A(\tau_i) = \int_{-\infty}^{\infty} \tilde{A}(\omega)e^{i\omega\tau_i}d\omega. \qquad (8)$$

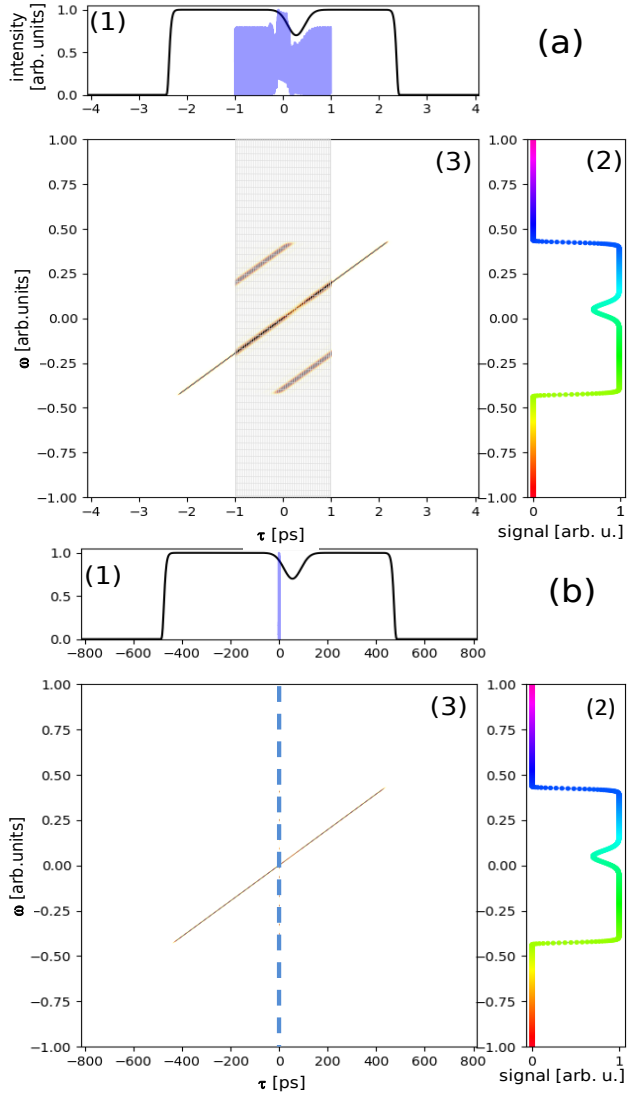

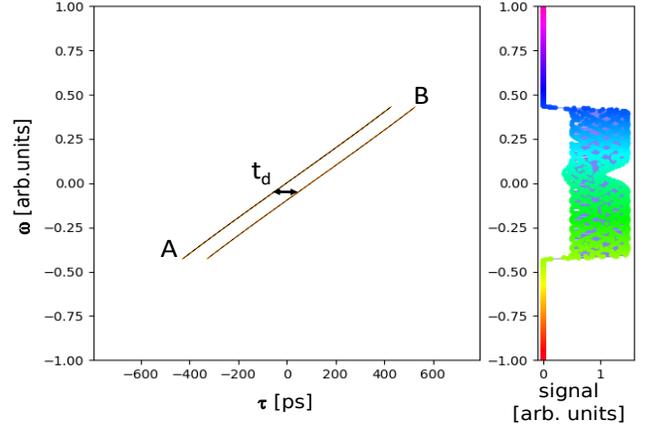

Fig. 3. Instantaneous frequency representation of the stretched main and post pulse. The main pulse (A) is followed by a post-pulse (B) with a delay of 100ps. The combination of two pulses results in a spectral interferometric pattern on the spectral axis.

Fig. 2. Instantaneous frequency representations of (a) a chirped pulse with moderate chirp, (b) a typical chirped pulse with stretching factor of $10^5$ typical for CPA lasers. In (a.3) in grey a reduced time window typically used for Fourier transformations is depicted together with the resulting aliasing. This leads to a distorted temporal representation (blue inset in (a.1). In (b.1) this temporal span is then even smaller for large stretching (blue line).

As expected by its linear nature, this operation is still done line-by-line (spectral amplitude) or column-by-column (temporal amplitude).

Combining the main pulse with a post-pulse, generated, e.g., by partial internal reflection in planar transmission optics, and stretching results in what occurs usually in CPA systems (Fig.3).

Without nonlinearities the pulses are then amplified with minor dispersions and some minor spectral narrowing due to amplification. With proper dispersion management, the main pulse is perfectly recompressed by the compressor and the output pulses look very similar to the input ones. PW-class CPA systems are designed to maximize the extracted laser pulse power, resulting in operation close to or within an intensity range of the stretched pulse that causes significant nonlinearities. This results in laser pulse degradation deleterious for ultra-high laser applications. The main effect is the Kerr effect globally characterized by the B-integral. This effect's most deleterious impact on the temporal contrast is the pre-pulse generation from a post pulse by temporal diffraction [6,10,11].

Let's consider the two pulses of Fig.3, $E_1$ representing the main pulse and $E_2$ a post-pulse time delayed by $t_d$. The third-order nonlinear polarization due to the Kerr effect [11],

$$P_{NL}(\tau) = \chi_3 [E_1(\omega_1(\tau)) + E_2(\omega_2(\tau))][E_1^*(-\omega_1(\tau)) + E_2^*(-\omega_2(\tau))][E_1(\omega_1(\tau)) + E_2(\omega_2(\tau))], \qquad (9)$$

gives rise to four wave mixing (FWM), self-phase modulation (SPM), cross-phase modulation (XPM) and cross-polarized wave generation (XPW). In particular, the terms $E_1(\omega_1(\tau))E_2^*(-\omega_2(\tau))E_1(\omega_1(\tau))$ and $E_2(\omega_2(\tau))E_1^*(-\omega_1(\tau))E_2(\omega_2(\tau))$ are the FWM processes of interest, giving the new frequencies $2\omega_1 - \omega_2$, and $2\omega_2 - \omega_1$. We see that a pre-pulse is generated $t_d$ before the main pulse and $\delta\omega_1 = (2\omega_1 - \omega_2)_\tau - \omega_1 = bt_d$, so it is constantly blue shifted from $\omega_1(\tau)$ as b>0 in PW-class CPA. Likewise-pulse is also generated $t_d$ after the post-pulse and $\delta\omega_2 = (2\omega_2 - \omega_1)_\tau - \omega_2 = -bt_d$ is red shifted from $\omega_2(\tau)$. Here $\delta\omega_1$ indicates its origin from the main pulse $\omega_1$ and $\delta\omega_2$ indicates its origin from the post-pulse. The FWM process is efficient as long as all spectral components are phase matched. The phase matching is kept since the components $\omega_1(\tau)$ and $\omega_2(\tau)$ are nearly equal. On the IFR, this temporal effect is simulated column by column. For any $\tau_i$, the temporal amplitude is calculated by using the inverse Fourier transform

$$\forall \tau_i, i \in \{1..N_\tau\}, A(t,\tau_i) = TF^{-1}[\tilde{A}(\omega,\tau_i)]. \qquad (10)$$

Then the nonlinear effect is applied in this time domain

iTEOX$$A_{out}(t,\tau_i) = |A_{in}(t,\tau_i)|^2 A_{in}(t,\tau_i). \quad (11)$$

And then the spectral amplitude at any $\tau_i$ is finally obtained by Fourier transform

$$\forall \tau_i, i \in \{1..N_\tau\}, \widetilde{A_{out}}(\omega,\tau_i) = TF[A_{out}(t,\tau_i)]. \quad (12)$$

The pre-pulse and post-pulse generation appears naturally form this processing as shown in fig.4.
If $\forall \tau_i, i \in \{1..N_\tau\}, b(\tau_i)$ is large enough then one can approximate: $A(t,\tau_i) \approx \left[\frac{\tilde{A}(\omega,\tau_i)}{b(\tau_i)}\right]$. Thus $\widetilde{A_{pre-pulse}}(\omega,\tau_i) \approx \alpha|\tilde{A}(\omega,\tau_i)|^2 \widetilde{A_{post-pulse}}(\omega,\tau_i)$, where $\alpha$ is the coupling factor from post-pulse to pre-pulse [12].

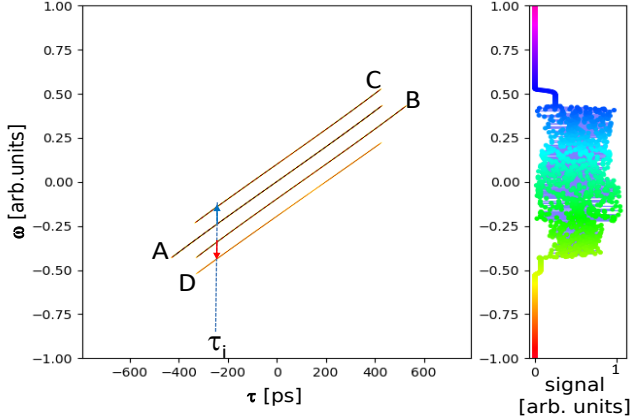

Fig. 4. Instantaneous frequency representation of stretched pulses after FWM due to the Kerr effect. The line labeled A is the main pulse, B is the post-pulse, C the generated pre-pulse and D the generated post-pulse. The spectral axis shows the interference in between the four different pulses.

The delay and frequency shift are directly represented as translations horizontally for the delay and vertically for the blue or red shifts. To facilitate the understanding and simplify the representation, pure chirp was used on this simulation, but higher dispersion orders must be used for any realistic CPA representation.

**II. Instantaneous Frequency representation for a realistic CPA laser simulation step-by-step.**

A typical CPA laser chain is depicted in Fig.5. We follow here the scheme presented by Liu et.al. [11]. The pulses are derived from a high repetition rate mode locked oscillator with an initial pulse length on the order of 25fs. The laser pulse train typically passes Pockels cells, Faraday isolators and polarizers to decrease the repetition rate and for back-reflection isolation within the laser chain. The repetition rate is reduced for amplification to the order of few Hz or less, especially in case of Petawatt class systems. After being stretched, typically from a time-bandwidth product of 0.5 to about $10^5$, or from 25fs up to 1ns, the laser pulse is amplified and re-compressed to close to the Fourier limit, determined by the spectral shape. The actively controlled spectral shape of the amplified pulse is typically a top hat, which is taken into account in all simulations presented here. Since the huge dispersion generated by the stretcher over-weights by far the dispersion of the material in the amplifier chain we will only consider the stretcher and compressor as dispersive elements for now.

Post-pulses are typically generated when the laser pulse passes through elements with plane-parallel surfaces. While for some cases this issue can be avoided by use of wedged components, causing spatio-temporal distortions degrading the laser pulse quality in the focus, for others it might not. Anti-reflection coatings reduce the relative pulse level of generated post-pulse to a certain degree. Plan-convex or plan-concave lenses can cause post-pulses within a certain angular acceptance as well. For the simulation discussed here we consider a rather large post-pulse level of 1% with respect to the main pulse and a delay of 100ps.

Taking all this into account the spectral group delay can be expressed as:

$$\begin{aligned}\tau(\omega) &= \sum_{n=2}^{\infty} \frac{\varphi_n}{(n-1)!}[(\omega-\omega_0)^{n-1}] \\ &\approx \varphi_2(\omega-\omega_0) + \frac{\varphi_3}{2}(\omega-\omega_0)^2 + \frac{\varphi_4}{3!}(\omega-\omega_0)^3\end{aligned} \quad (13)$$

where $\varphi_2$ is the linear chirp mainly stretching the pulse called group velocity dispersion (GVD), $\varphi_3$ is the third order spectral phase (third order dispersion TOD), i.e. the first distortion order on the linear chirp, and $\varphi_4$ is the fourth order spectral phase (fourth order dispersion FOD), i.e. the second distortion order on the linear chirp. As sketched in Fig.5.b, these distortions modify the previously used simplified IFR where only linear chirp was applied, also in contrast to Liu et.al.[11].

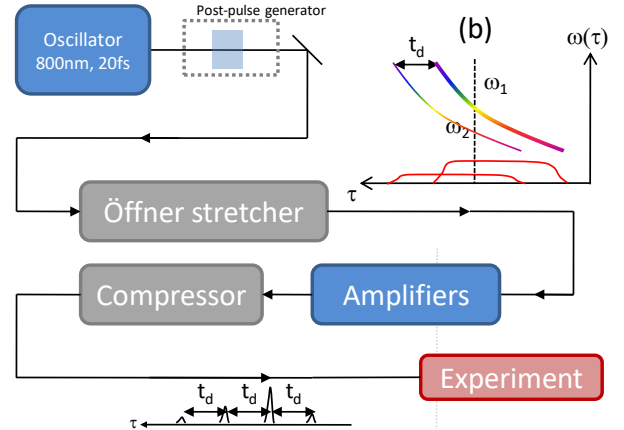

Fig.5. Schematic diagram of a typical laser chain including post-pulse generation.

The pulse and its post-pulse will undergo the Kerr effect induced FWM dominantly induced by the amplifier materials [10]. After amplification the pulses are recompressed by the compressor such that the main pulse is near-perfectly dispersion compensated and compressed as in the previous example.

The first simulation considers only pure linear chirp stretching (Fig.6). The overall B-integral is chosen such that the pre-pulse generated yields about the same power as the post-pulse [12], meaning about $\sqrt{3}$. As expected the Kerr effect through FWM generates in the output pulse a "time diffracted" pre-pulse blue shifted in frequency. A post-pulse should also appear but it is too weak to be visible on the dynamic scale of 80dB. In the ideal case of a pure chirp, as already mentioned by Liu et al. [11], the generated and compressed pre-pulse appears exactly with the mirrored delay of the original post pulse, but with significant frequency shifts (Figs.6.a-c). This blue shift can be significant compare to the pulse bandwidth and can lead to an underestimated temporal deterioration by Third Order Cross-Correlators with limited bandwidth [13]. The generated post-pulse can also be resolved on the temporal intensity display.

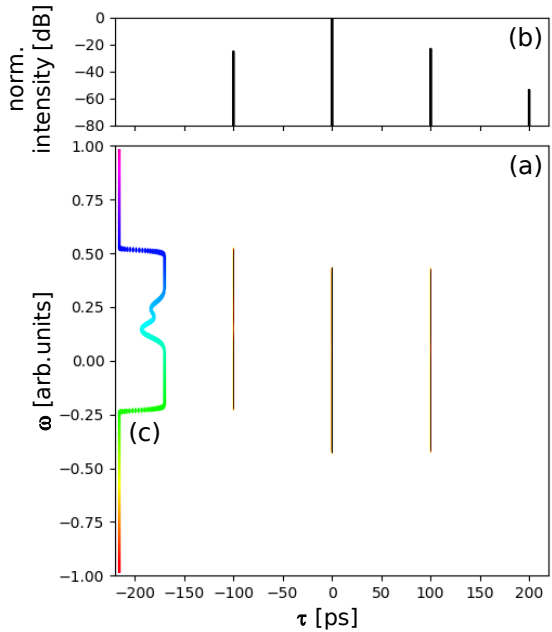

Fig. 6. (a) Instantaneous frequency representation after compression of a linear chirp, (b) temporal intensity on log scale (dB), (c) normalized pre-pulse spectrum.

The simulation was performed on a standard personal computer with a 2000 points array in few seconds without any code optimization. The chirp used stretches the pulse by a factor of 40000 (1ns stretched pulse) so the delay of 100ps is still observable (Fig.3). Smaller or larger delays could be used without significant impact on simulation time and precision of the result.

In the more interesting case of distortions of the linear chirp, commonly present in CPA stretcher and compressor realizations, the generated pre-pulse is modified and not recompressed completely at the output by the compressor. In case of third order spectral phase distortion, the pre-pulse is chirped (Fig.7) and its peak power can be decreased by two orders of magnitude. In contrast, the generated post-pulse appears to be not affected, changing the relative pulse height of pre- and post-pulse. The apparent delay of the pre-pulse is also shifted to a smaller delay than the equivalent delay of the post-pulse by symmetry.

For a fourth order spectral phase distortion, the pre-pulse exhibits a third order spectral phase distortion (Fig.8). Its maximum peak power is also decreased but less than for third order distortion. The generated post pulse has disappeared below -80dB. In both cases, interestingly, the peak power of the pre-pulse is decreased by this stretching detuning from pure linear chirp.

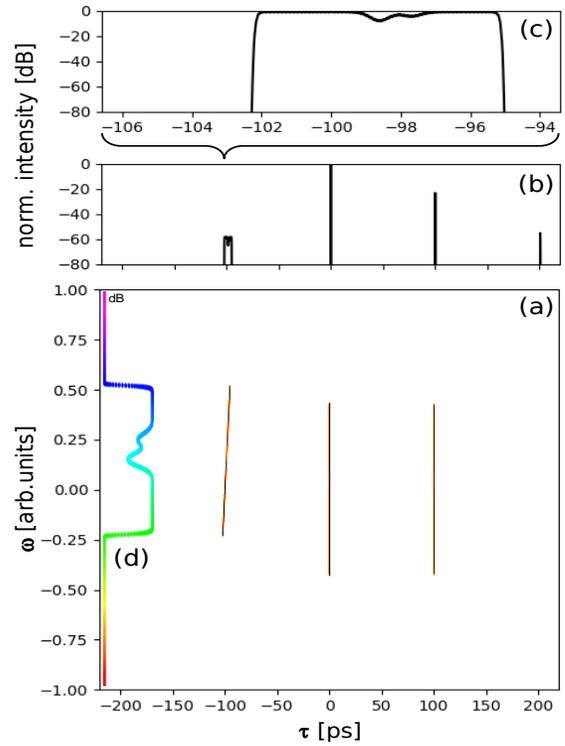

Fig. 7. (a) Instantaneous frequency representation with third order spectral phase distortion in stretching, (b) its temporal intensity on log scale (dB), (c) zoom of temporal intensity of the pre-pulse on log scale (dB), (d) normalized pre-pulse spectrum.

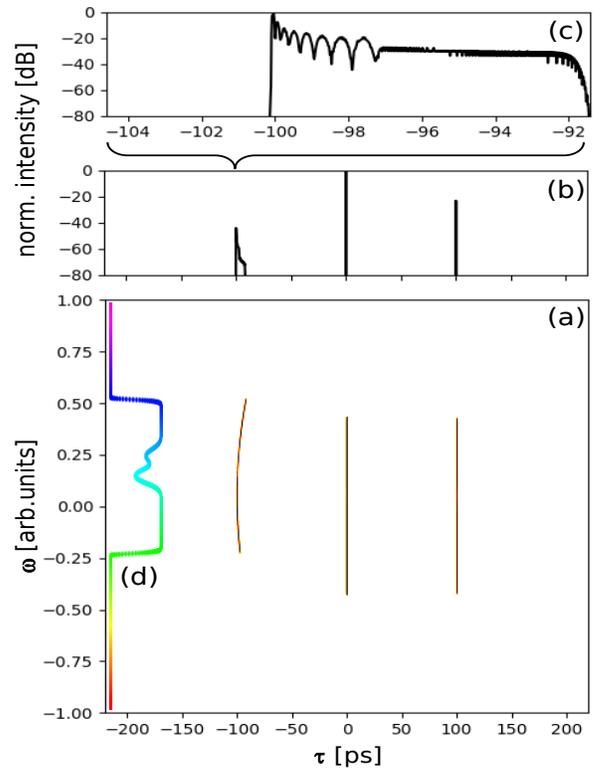

Fig. 8. (a) Instantaneous frequency representation with fourth order spectral phase distortion in stretching, (b) its temporal intensity on log scale (dB), (c) zoom of temporal intensity of the pre-pulse on log scale (dB), (d) normalized pre-pulse spectrum.

This pre-pulse temporal intensity distortion is similar to the one observed by Kiriyama et al. [6]. A proper stretching distortion could thus be intentionally applied to decrease the pre-pulse peak power and enhance the temporal contrast of the laser. This observation can be understood by the effect of the blue shift on the compression. Without blue shift, the compressor fully compensates the dispersion of the system. With the blue shift, the compensation is incomplete.:

$$\begin{aligned}\delta\tau_{\delta\omega_1}(\omega) &= \sum_{n=2}^{\infty}\frac{\varphi_n}{(n-1)!}\left[\begin{array}{c}(\omega-\omega_0)^{n-1}\\-(\omega+\delta\omega_1-\omega_0)^{n-1}\end{array}\right]\\ &= \sum_{n=2}^{\infty}\frac{\varphi_n}{(n-1)!}(-\delta\omega_1)\left[\sum_{k=0}^{n-2}\sum_{j=0}^{k}\binom{k}{j}(\delta\omega_1)^j(\omega-\omega_0)^k\right]\end{aligned} \quad (14)$$

If we consider the effect nearby $\omega_0$, $\omega - \omega_0 \ll \delta\omega_1$ and that $\delta\omega_1$ is nearly constant over the bandwidth, the Nth order distortion for the pre-pulse is produced by a combination of higher order terms of the stretcher dispersion:

$$\varphi_{M,prepulse}(\omega_0) = -\sum_{n=M+1}^{\infty}\frac{\varphi_n}{(M-1)!(n-M)!}(\delta\omega_1)^{n-M} \quad (15)$$

For pure third order distortion, the chirp of the pre-pulse can be approximated to

$$\frac{\partial\delta\tau_{\delta\omega_1}(\omega)}{\partial\omega} \approx -[(\varphi_3\delta\omega_1)]. \quad (16)$$

For pure fourth order, the chirp is combined with a third order

$$\frac{\partial\delta\tau_{\delta\omega_1}(\omega)}{\partial\omega} \approx -\left[\left(\frac{\varphi_4}{2!}\delta\omega_1^2\right)\right] - \left[\frac{\varphi_4}{2!}\delta\omega_1\right][(\omega-\omega_0)]. \quad (17)$$

These approximations confirm the effects seen on fig.7 and fig.8. Non-intuitively, having a large third order may ease the constraints on the pulse contrast as it decreases the peak power of the generated pre-pulse due to its residual chirp.

Note, however, that this model lacks precision because it assumes that $\delta\omega_1 = (\omega_1-\omega_2)_\tau$ is constant.

This realistic example of a case currently limiting high-intensity laser performance on target illustrates the main advantage of the presented IFR simulation method. It is compatible with huge time-bandwidth product pulses existing in most high power CPA laser systems. The temporal domain can be stretched and recompressed without loss of information and without any modification on the frequency domain. It is also very efficient numerically as all the operations presented above are purely along one dimension.

## VI. Conclusion

We have introduced a novel method for the simulation of nonlinearities on very large time-bandwidth product pulses. This method is very efficient numerically, as well as intuitive for the interpretation of the physical meaning of the pulse modifications. It has been applied to model pre-pulse generation in CPA chains as an example of current interest and is planned to be used for a quantitative description of a full laser chain. Based on the investigated principles it could be shown already in the simplified picture of single higher-order stretching distortions that a contrast enhancement, i.e. a reduction of the intensity of the generated pre-pulse, becomes possible due to its corresponding incomplete compression.